\documentclass[a4paper,11pt]{article}
\usepackage[ams]{}
\usepackage{amssymb}
\usepackage{graphicx}
\usepackage[theorem]{}
\begin{document}
\title{A Jordan GNS Construction for the Holonomy-Flux *-algebra}
\author{Michael Rios\footnote{email: mrios4@calstatela.edu}\\\\\emph{California State University, Los Angeles}\\\emph{Mathematics Graduate Program}
\\\emph{5151 State University Drive}\\\emph{Los Angeles, CA 90032-8531}  } \date{\today}\maketitle
\begin{abstract}
The holonomy-flux *-algebra was recently proposed as an algebra of basic kinematical observables for loop quantum gravity.  We show the conventional GNS construction breaks down when the the holonomy-flux *-algebra is allowed to be a Jordan algebra of observables.  To remedy this, we give a Jordan GNS construction for the holonomy-flux *-algebra that is based on trace.  This is accomplished by assuming the holonomy-flux *-algebra is an algebra of observables that is also a Banach algebra, hence a JB algebra.  We show the Jordan GNS construction produces a state that is invariant under all inner derivations of the holonomy-flux *-algebra.  Implications for the corresponding Jordan-Schr$\ddot{\textrm{o}}$dinger equation are also discussed.\\\\
$Keywords:$ Jordan GNS construction, holonomy-flux *-algebra.
\end{abstract}
\newpage
\tableofcontents
\section{Introduction}
A formal quantization of loop quantum gravity was recently carried out in \cite{1}, where the kinematical algebra was defined as a *-algebra of observables, called the holonomy-flux *-algebra.  The problem of uniqueness of the Hilbert space representation was addressed through the search for a state that is invariant under automorphisms of the holonomy-flux *-algebra.\\
\indent Here we address the representation problem by showing the conventional GNS construction fails when the holonomy-flux *-algebra is allowed to be a Jordan algebra of observables.  It has been shown that the proper GNS construction for observable algebras is based on trace \cite{4}.  We follow through the Jordan GNS construction for the holonomy-flux *-algebra and demonstrate that trace vanishes for all inner derivations of the holonomy-flux *-algebra.  Following \cite{4}, we lastly give the corresponding Jordan-Schr$\ddot{\textrm{o}}$dinger equation for the resulting quantum mechanical system.  
\section{JB Algebras and Jordan $C^*$-algebras}
Following \cite{2,3}, we review properties of JB algebras and Jordan $C^*$-algebras.
\subsection{JB Algebras}
\textbf{Definition 2.1.1}  A \textit{Jordan algebra} is a real vector space \textit{A} equipped with the Jordan product (i.e.  a bilinear form) $(a,b)\rightarrow a\circ b$ satisfying $\forall a,b\in A$:
\begin{center}$a\circ b = b\circ b$,\end{center}
\begin{center}$a\circ (b\circ a^2) = (a\circ b)\circ a^2$. \end{center}
\textit{A} is \textit{unital} if it admits a unit with respect to the Jordan product.
\textit{Inner derivations} of the Jordan algebra are given through the associator:\\
\begin{center}$[r,a,s]=(r\circ a) \circ s - r\circ (a \circ s),$\end{center}
which acts as a derivation on its second argument for self-adjoint $r,s\in A$.\\\\
\textbf{Definition 2.1.2}  A \textit{JB algebra} is a Jordan algebra that is also a Banach algebra with respect to the Jordan product $\circ$ with a norm $||\cdot||$ that satisfies $\forall a,b\in A$:
\begin{center}$||a^2||\leq ||a^2+b^2||$,\end{center}
\begin{center}$||a^2||=||a||^2$.\\ \end{center}
If the JB algebra is the dual of a Banach space, it is called a \textit{JBW algebra}.\\\\
\textbf{Example 2.1.1}  The JB algebra of self-adjoint operators acting on a Hilbert space $\mathcal{H}$ is endowed with the Poisson-like product $a \circ b = \frac{1}{2}(ab+ba)$.
\subsection{Jordan $C^*$-algebras}
\textbf{Definition 2.2.1}  Let $\mathcal{A}$ be a complex Banach space and a complex Jordan algebra equipped with an involution *.  Then $\mathcal{A}$ is a \textit{Jordan} $C^*$\textit{-algebra} if the following conditions hold $\forall x,y,z\in\mathcal{A}$:
\begin{center}$||x\circ y||\leq ||x||||y||$\\ \end{center}
\begin{center}$||z||=||z^*||$\\ \end{center}
\begin{center}$||\{zz^*z\}||=||z||^3$.\\ \end{center}
\textbf{Theorem 2.2.1 (Wright)}\\
Each JB algebra is the self-adjoint part of  a unique Jordan $C^*$-algebra.
\section{The Jordan GNS Construction}
In this section we follow through a Jordan GNS construction \cite{4} for the holonomy-flux *-algebra.  The Jordan GNS construction makes use of a Jordan algebra of observables instead of an associative $C^*$-algebra, resulting in a Hilbert space built from observables.
\subsection{The Representation Problem for Observable Algebras}
\indent Let $\mathfrak{A}$ be a Jordan *-algebra of observables.  Following \cite{1}, define
\begin{equation}
\mathfrak{J}=\{a\in\mathfrak{A}\hspace{.1cm}|\hspace{.1cm}\omega(a^*a)=0\}.
\end{equation}
As $\mathfrak{A}$ is a Jordan *-algebra of observables, we explicitly re-express this as
\begin{equation}
\mathfrak{J}=\{a\in\mathfrak{A}\hspace{.1cm}|\hspace{.1cm}\omega(a^*\circ a)=\omega(a^2)=0\}.
\end{equation}
$\mathfrak{J}$ is not an ideal in $\mathfrak{A}$, so $\mathfrak{A}$ does not act on the Hilbert space:
\begin{equation}
\mathcal{H}_\omega:=\overline{\mathfrak{A}/\mathfrak{J}}.
\end{equation}
It follows that the conventional GNS construction is not suitable for Jordan algebras of observables.  In \cite{4}, the Jordan GNS construction based on trace was given to allow a Hilbert space representation for algebras of observables.
\subsection{The Hilbert Space Construction}
\indent To proceed with the Jordan GNS construction, the holonomy-flux *-algebra $\mathfrak{A}$ is defined as a Jordan Banach *-algebra of observables, a JB algebra.  Additionally, we let $\mathfrak{A}$ be a JBW algebra, so there always exists a semi-finite, faithful, normal trace on $\mathfrak{A}$ \cite{5,6}. $\mathfrak{A}^+_1$ consists of elements of $\mathfrak{A}$ with finite trace and $\mathfrak{A}_2$ are those $a\in\mathfrak{A}$ for which $tr (a^*\circ a)<\infty$. The trace $tr$ induces a bilinear, symmetric, real inner product on $\mathfrak{A}_2$ as
\begin{equation}
<a|b>=tr (a\circ b).
\end{equation}
It is positive-definite and yields a norm
\begin{equation}
||a||_{tr}=\sqrt{tr\hspace{.1cm}a^2}.
\end{equation}
Closure with respect to the norm yields a Hilbert space
\begin{equation}
\mathcal{H}_{tr}:=\overline{\mathfrak{A}_2}^{tr}
\end{equation}
which is acceptable as $\mathfrak{A}_2$ is an ideal.
\subsection{The Hilbert Space Representation} 
The representation of $\mathfrak{A}$ on the Hilbert space $\mathcal{H}_{tr}$ proceeds with the use of a Jordan module \cite{4}.  Let $V$ be a vector space and define two bilinear mappings $(\mathfrak{A},V)\rightarrow V:(a,v)\mapsto a.v$ and $(\mathfrak{A},V)\rightarrow V:(a,v)\mapsto v.a$. $V$ is a Jordan module if for any $v\in V$ and $a,b\in\mathfrak{A}$
\begin{displaymath}
a.v=v.a
\end{displaymath}
\begin{displaymath}
a^2.(a.v)=a.(a^2.v)
\end{displaymath}
\begin{displaymath}
2a.(b.(a.v))+(b\circ a^2).v=2(a\circ b).(a.v)+a^2.(b.v)
\end{displaymath}
A Jordan representation is defined by the linear mapping
\begin{displaymath}
\pi_J:\mathfrak{A}\rightarrow Hom_R(V,V)
\end{displaymath}
satisfying
\begin{displaymath}
\pi_J(a^2)\pi_J(a)=\pi_J(a)\pi_J(a^2),
\end{displaymath}
\begin{displaymath}
2\pi_J(a)\pi_J(b)\pi_J(a)+\pi_J(b\circ a^2)=2\pi_J(a\circ b)\pi_J(a)+\pi_J(a^2)\pi_J(b).
\end{displaymath}
Let $b\in\mathfrak{A}_2$.  For $a\in\mathfrak{A}$ there exists a multiplication operator $T_a:\mathfrak{A}_2\rightarrow\mathfrak{A}_2$ defined by
\begin{displaymath}
T_a b=a\circ b
\end{displaymath}
$\mathfrak{A}_2$ is a Jordan module under the multiplication operator $T_a$.  By continuity, $T_a$ can be extended to all of $\mathcal{H}_{tr}$.  The expectation value, in density matrix form $(||b||_{tr}=1)$, is now given by
\begin{equation}
\omega_b(a)=<b|a|b>=tr (b^2\circ a)=tr(\rho\circ a) .
\end{equation}
\subsection{The Jordan-Schr$\ddot{\textrm{o}}$dinger Equation}
It is easily seen that trace vanishes for all inner derivations of $\mathfrak{A}$ using the properties $tr\hspace{.1cm}(a\circ b)\circ c=tr\hspace{.1cm} a\circ (b\circ c)$ and $tr(a+b)=tr\hspace{.1cm}a + tr\hspace{.1cm}b$, giving:
\begin{equation}
tr\hspace{.1cm}[r,a,s]=0.
\end{equation}
This has clear implications for the resulting Jordan-Schr$\ddot{\textrm{o}}$dinger equation \cite{4}
\begin{equation}
\dot{v}=[r,v,s].
\end{equation}
For by taking the trace of the Jordan-Schr$\ddot{\textrm{o}}$dinger equation we have
\begin{equation}
tr\hspace{.1cm}\dot{v}=tr\hspace{.1cm}[r,v,s]=0
\end{equation}
Trace is thus invariant with respect to the quantum evolution equation.
\section{Conclusion}
We showed that when the holonomy-flux *-algebra is a Jordan algebra, the conventional GNS construction breaks down.  Following \cite{4}, we gave a Jordan GNS construction for a JBW holonomy-flux *-algebra.  The Jordan GNS construction produced a Hilbert space involving observables only.  This is in accordance with \cite{1}, where the holonomy-flux *-algebra is an algebra of observables, rather than a general $C^*$-algebra.  The difference in construction is that state takes the form of trace in the JB formalism.  This allowed us to show that trace vanishes for all inner derivations of the JBW holonomy-flux *-algebra, giving the Jordan analog of the Yang-Mills gauge invariant and diffeomorphism invariant state defined in \cite{1}.\\
\indent The JB formalism further led to a Jordan-Schr$\ddot{\textrm{o}}$dinger equation for the holonomy-flux *-algebra, expressed in terms of inner derivations given by the Jordan associator.  Trace was shown to be invariant with respect to this Jordan-Schr$\ddot{\textrm{o}}$dinger equation.


\begin{thebibliography}{-label}
\bibitem[1]{1}J. Lewandowski, A. Okolow, H. Sahlmann, T. Thiemann, \textit{Uniqueness of diffeomorphism invariant states on holonomy-flux algebras}, \texttt{gr-qc/0504147}.
\bibitem[2]{2}J. Hamhalter, \textit{Universal State Space Embeddability of Jordan Banach Algebras}, Proc. of AMS. 127 (1999), 131-137.
\bibitem[3]{3}J. D. M. Wright, \textit{Jordan $C^*$-algebras}, Mich. Math. J. 24 (1977), 291-302.
\bibitem[4]{4}W. Bischoff, \textit{On a Jordan-algebraic formulation of quantum mechanics: Hilbert space construction}, \texttt{hep-th/9304124}.
\bibitem[5]{5}O. Bratteli, D. W. Robinson, \textit{Operator Algebras and Quantum Stat. Mech. 1}, (Springer, New York, 1987).
\bibitem[6]{6}M. Takesaki, \textit{Theory of Operator Algebras I}, (Springer, New York, 1979).
\bibitem[7]{7}S. Catto, \textit{Exceptional Projective Geometries and Internal Symmetries}, \texttt{hep-th/0302079}.
\end{thebibliography}
\end{document}